\documentclass[useAMS,usenatbib,useasmath]{mnras}
\usepackage[dvips]{graphicx}
\usepackage{longtable}
\usepackage{times}
\usepackage{amsmath}
\usepackage{ulem}
\usepackage{subfig}
\usepackage{color}
\usepackage{xcolor}
\usepackage{hyperref}
\hypersetup{
	colorlinks = true,
	urlcolor = blue,
	linkcolor = blue,
	citecolor = blue
}

\def\pasp{PASP}
\def\apj{ApJ}
\def\apjs{ApJS}
\def\apjl{ApJL}
\def\aap{A\&A}

\def\aaps{A\&AS}
\def\mnras{MNRAS}
\def\nat{Nature}
\def\iaucirc{IAU~Circ.}

\newcommand{\Msun}{\hbox{$M_\odot$}}
\newcommand{\Mch}{\hbox{M$_{\rm Ch}$}}

\newcommand{\hi}{\mbox{H~{\sc i}}}

\newcommand{\kms}{\hbox{km$\,$s$^{-1}$}}
\newcommand{\ergsec}{\hbox{ergs$\,$s$^{-1}$}}

\newcommand{\iso}[2]{$^{#2}{\rm #1}$}

\def\lesssim{\mathrel{\hbox{\rlap{\hbox{\lower4pt\hbox{$\sim$}}}\hbox{$<$}}}}
\def\gtrsim{\mathrel{\hbox{\rlap{\hbox{\lower4pt\hbox{$\sim$}}}\hbox{$>$}}}}
\def\gray{$\gamma$-ray}
\def\grays{$\gamma$-rays}

\newcommand{\CMFGEN}{{\sc cmfgen}}
\def\fig{Fig.}
\def\figs{Figs.}
\def\eq{Eq.}
\def\eqs{Eqs.}
\def\rrte{relativistic radiative transfer equation}

\title[Solving $\gamma$-ray rad. transfer eqn. for SNe]
{Solving the gamma-ray radiative transfer equation for supernovae}
\author[Kevin D. Wilk, D. John Hillier, Luc Dessart]
{Kevin D. Wilk$^{1}$\thanks{E-mail: kdw25@pitt.edu}, D. John Hillier$^1$\thanks{E-mail: hillier@pitt.edu}, Luc Dessart$^2$ \\
$^1$ Department of Physics and Astronomy \& Pittsburgh Particle physics, Astrophysics, and Cosmology Center (PITT PACC), University of Pittsburgh,  \\ Pittsburgh, PA 15260, USA \\
$^2$ Unidad Mixta Internacional Franco-Chilena de Astronom\'ia (CNRS UMI 3386), 
Departamento de Astronom\'ia, Universidad de Chile, \\
Camino El Observatorio 1515, Las Condes, Santiago, Chile}
\begin{document}

\date{Accepted . Received }

\pagerange{\pageref{firstpage}--\pageref{lastpage}} \pubyear{2016}

\maketitle

\label{firstpage}

\begin{abstract}
We present a new relativistic radiative-transfer code for \grays\ of energy less than 5 MeV in supernova (SN) ejecta. This code computes the opacities, the prompt emissivity (i.e. decay), and the scattering emissivity, and solves for the intensity in the co-moving frame. 
Because of the large expansion velocities of SN ejecta, we ignore redistribution effects associated with thermal motions. The energy deposition is calculated from the energy removed from the radiation field by scattering or photoelectric absorption.
This new code yields comparable results to an independent Monte Carlo code. However, both yield non-trivial differences with the results from a pure absorption treatment of \gray\ transport. 
A synthetic observer's frame spectrum is also produced from the CMF intensity. At early times when the optical depth to \grays\ is large, the synthetic spectrum show asymmetric line profiles with redshifted absorption as seen in SN 2014J. 
This new code is integrated within {\sc cmfgen} and allows for an accurate and fast computation of the decay energy deposition in SN ejecta.
\end{abstract}

\section{Introduction}
Supernovae (SNe) are luminous astrophysical events, and studies of SNe probe stellar evolution of the progenitor and reveal properties of the explosion mechanism. Understanding both spectra and light curves allows us to investigate the physics and retrieve SN ejecta properties. For instance, Type Ia SNe  produce large amounts of radioactive material that control the thermal evolution of the ejecta by non-thermal heating. 

For decades, the standard paradigm has been that SNe arise through two mechanisms: gravitational core-collapse (CC) and thermonuclear (Type Ia). Historically, SNe have been classified into two spectral types, Type I (no \hi\ lines) and Type II (strong \hi\ lines) \citep{Minkowski1941}. However, the most successful theory of SNe is that  SNe of Type Ib, Ic, and Ibc, and II result from core-collapse -- see \cite{Colgate1966,Burrows1995,Janka1996,Mezzacappa1998}. On the other hand, Type Ia SNe are believed to be thermonuclear explosions of carbon-oxygen (CO) white dwarfs (WDs) \citep{Hoyle1960}. These thermonuclear explosions produce large amounts of radioactive material ($\sim$0.6 \Msun) \citep[see, e.g.,][]{Scalzo2014b}, mainly \iso{Ni}{56}, which decays into \iso{Co}{56} and then \iso{Fe}{56}. Core-collapse SNe (CCSNe) are thought to produce about an order of magnitude less \iso{Ni}{56} than this ($\sim$$10^{-2}$--$10^{-3}$ \Msun), which is what powers the late time light curve \citep[see review of CCSNe]{Janka2012}. 

A crucial issue for modelling SNe is the location of radioactive material. If the radioactive material is mixed into the outer ejecta, it will heat it and enhance the ionization. In Type
Ia SNe, mixing of \iso{Ni}{56} has been invoked to explain the brightness and colour at very early times \citep{Hoeflich1998,Hoeflich2002,Woosley2007,Hoeflich2017}. It was also invoked in SN1987A to explain the early detection of X-rays and \grays\ from 1987A \citep[][and references therein]{PintoWoosley1988Nature,PintoWoosley1988ApJ,Bussard1989,TBB1990,DessartHillierLi2012}.

The peculiar Type II SN, SN1987A, is the only CCSN for which we have detected the \iso{Co}{56} decay lines at 847 and 1238 keV \citep{Makino1987,Matz1988a,Matz1988b,Matz1988c,Matz1988Nature,Sunyaev1987Nature,Tanaka1988,Cook1988}. After SN1987A was observed, models of expected late time (1--2 years post-explosion due to high initial column densities) \gray\ and X-ray fluxes and profile shapes calculated from Monte Carlo radiative transfer soon followed \citep[][and references therein]{PintoWoosley1988Nature,PintoWoosley1988ApJ,Bussard1989,TBB1990}.  To date, SN 2014J is the only Type Ia SN with \gray\ detections \citep{Churazov2014,Churazov2015}. 

Many \gray\ radiative-transfer codes have utilized Monte Carlo techniques to treat the radiative transfer \citep{Pozdnyakov1983,Hoeflich1992,Milne2004,Sim2007,SimMazzali2008,Hillier2012,Summa2013}. Another technique, used by \cite{Swartz1995} and \cite{Jeffery1998}, utilizes a grey transfer approach to treat \gray\ transport. \cite{Swartz1995} finds that the value of $\kappa_\gamma=0.06Y_e$ cm$^2$ g$^{-1}$, where $Y_e$ is the total number of electrons per baryon, best describes the interaction of \grays\ in the SN ejecta. In contrast, the work presented here is the first of its kind to formally solve the radiative transfer equation for \grays\ for SN ejecta. 

This paper is organized as follows: In Section~\ref{tech} we outline the method used to calculate the opacity (Compton scattering and X-ray photoelectric absorption) and emissivity (prompt emission and scattering) needed to solve the \rrte. The implementation of our method into \CMFGEN\ is discussed in Section~\ref{implem}. In Section~\ref{results} we illustrate our results using a SN Ia ejecta resulting from a delayed-detonation in a Chandrasekhar mass (\Mch) WD from \cite{Wilk2018}. We also present synthetic \gray/X-ray spectra around bolometric maximum and at nebular times, and compare our results with those from a Monte Carlo calculation and those obtained using the grey approximation. Finally, in Section~\ref{conclusion}, we summarise our results.
%
%
\section{Technique}\label{tech}
%
%
We developed this code for implementation into \CMFGEN\ \citep{Hillier1998,Hillier2012,Dessart2014b}, which is a radiative transfer code that solves the spherically symmetric, non-local-thermodynamic-equilibrium (non-LTE), time-dependent, \rrte\ in the co-moving frame (CMF). This work was undertaken as a consistency check of the Monte Carlo (MC) radiative transfer code utilized by \CMFGEN\ \citep{Hillier2012}, and to provide an alternative technique to track photons and subsequent Compton scatterings or photon absorption for computation of the energy deposition in SNe. Since the expansion velocities dominate over thermal motions, this work ignores effects of thermal redistribution. 
%
%
\subsection{Radiative Transfer Equation}\label{rte_section}
%
%
We implement the code by solving the \rrte\ along rays as outlined in \cite{OlsonKunasz1987}, \cite{Hauschildt1992}, and \cite{Hillier2012}. The \rrte\ is 
\begin{eqnarray}\label{rrte_eqn}
	\frac{\gamma(1+\beta\mu)}{c}\frac{\partial I_\nu}{\partial t}\,+\,\gamma(\mu+\beta)\frac{\partial I_\nu}{\partial r} \nonumber \\
    +\, \gamma(1-\mu^2)\left[\frac{1+\beta\mu}{r}\,-\, \Lambda \right]\frac{\partial I_\nu}{\partial \mu} \nonumber\\
   -\, \gamma\nu \left[\frac{\beta(1-\mu^2)}{r}\,+\,\mu\Lambda\right]\frac{\partial I_\nu}{\partial \nu} \nonumber \\
   +\, 3\gamma\left[\frac{\beta(1-\mu^2)}{r}\,+\,\mu\Lambda\right]I_\nu &=\; \eta_\nu - \chi_\nu I_\nu,
\end{eqnarray}
where $\beta=v/c$, $\;\gamma=1/\sqrt{1-\beta^2}$, $\;\mu=\cos\theta$, and 
\begin{equation}
	\Lambda = \frac{\gamma^2(1+\beta\mu)}{c}\frac{\partial \beta}{\partial t} + \gamma^2(\mu+\beta)\frac{\partial \beta}{\partial r}.
\end{equation}

In \eq~\ref{rrte_eqn}, the specific intensity, emissivity, and opacity (all measured in the CMF) are assumed to be functions of several variables $[I_\nu$=$I(t,\, r,\, \mu,\, \nu), \; \eta_\nu$=$\eta(t,\, r,\, \mu,\, \nu), \text{ and }\chi_\nu$=$\chi(t,\, r,\, \nu)]$. However, if we ignore all time dependence, we can reduce this equation along characteristic rays reducing the partial differential equation with dependent variables $(r,\, \nu,\, \mu)$ to a partial differential equation with dependent variables $(s,\nu)$ \citep{Mihalas1980}. Time dependence can be neglected as \grays\ undergo few scatterings before the energy is deposited or the photon escapes to the observer. 

From \eq~\ref{rrte_eqn}, our characteristic equations are
\begin{equation}
	\frac{dr}{ds} = \gamma(\mu\,+\,\beta)\hspace{0.3cm}\text{and}\hspace{0.3cm}\frac{d\mu}{ds}=\gamma(1-\mu^2)\left[\frac{1+\beta\mu}{r}\,-\, \Lambda \right].
\end{equation}

We can now write the \rrte\ along a characteristic ray as
\begin{equation}\label{rrte_char_eqn}
	\frac{\partial I_\nu}{\partial s}\,-\,\nu\Pi\frac{\partial I_\nu}{\partial\nu}\,=\, \eta_\nu \,-\, (\chi_\nu \,+\, 3\Pi) I_\nu,
\end{equation}
where 
\begin{equation}
	\Pi \,=\, \gamma\left[\frac{\beta(1-\mu^2)}{r} \,+\,\mu\Lambda \right]
\end{equation}
In order to solve \eq~\ref{rrte_char_eqn}, we use a backward differencing in frequency (i.e. $\partial\nu=\nu_{i-1}-\nu_i$ with $i$ denoting the current frequency). We then solve \eq~\ref{rrte_char_eqn} by usual means for the formal solution along each ray for each frequency.
%
%
\subsection{Opacities}\label{opacity_section}
%
%
Most nuclear decay lines in SNe have energies less than 3.5 MeV (see Table~\ref{decay_data}). For energies less than this, the dominant opacity source is Compton scattering and photoelectric absorption. Below 100 keV, the dominant opacity is photoelectric absorption and above that it is Compton scattering -- \cite[see figure 1 in][]{Milne2004}. This work only incorporates both photoelectric absorption and Compton scattering opacity. We neglect the influence of $e^-  e^+$ pair production opacity because the typical decay \gray\ energies are less than 3.5 MeV in SNe. 

For comparison with the MC method of \cite{Hillier2012} which follows that of \cite{Kasen2006}, we use a photoelectric absorption opacity given by
\begin{equation}\label{xray_opac_eqn}
	\chi_{\nu}^{\rm abs} = \left(\frac{m_e c^2}{h\nu}\right)^{3.5} \sigma_{\rm T} \alpha^{4}8\sqrt{2} \sum_{i}^{N_{\rm spec.}} N_{i} Z_i^{5},
\end{equation}
where $m_e$ is the electron mass, $\sigma_{\rm T}$ is the Thomson cross section, $\alpha$ is the fine structure constant, $N_i$ is the number density of species $i$, and $Z_i$ is the atomic number of species $i$.
The Compton scattering opacity as given by eq.~7.113 of  \cite{Pomraning1973} is
\begin{eqnarray}\label{compton_opac_eqn}
	\chi_{\nu}^{\rm C} = 2\pi r^2_{e}N_{e} \left[\left( \frac{1+x}{x^3}\right)\left\{\frac{2x(1+x)}{1+2x}-\log(1+2x) \right\} + \right. \nonumber \\ 
    \left. \frac{\log(1+2x)}{2x} - \frac{1+3x}{(1+2x)^2} \right],
\end{eqnarray}
where $N_{e}$ is the number density of electrons, $r_{e}$ is the classical electron radius, and $x$ is $h\nu/m_{e}c^2$. 

%
%
\subsection{Emissivities}\label{emissivity_section}
%
%
The total emissivity in the \rrte\ has two components. The first component is an isotropic prompt emission from nuclear decays, and the second is the scattering emissivity arising from Compton scattering.
\subsubsection{Prompt Emission}
The simpler of the two, the isotropic emissivity from the prompt decays is given by
\begin{equation}\label{iso_emiss_eqn}
		\eta_\nu^{\rm iso} = \frac{1}{4\pi} \sum_{i=1}^{N_{\rm isot.}}\sum_{j=1}^{N_{\rm lines}}\frac{N_{i}}{\tau_i}E_{ij}P_{ij}\frac{e^{-\Gamma}}{\sqrt{2\pi}V_{\rm Dop}\nu_{ij}/c},
\end{equation}
where $\Gamma=\frac{1}{2}[(\nu-\nu_{ij})/(V_{\rm Dop}\nu_{ij}/c)]^2$, $N_i$ is the number density of the $i$-th species isotope, $\tau_i=(t_{1/2})_i/\ln(2)$ is the nuclear decay constant for the $i$-th species isotope (see Table~\ref{decay_data} for half-lives -- $t_{1/2}$ -- of \iso{Ni}{56} and \iso{Co}{56}), $E_{ij}(\nu_{ij})$ is $j$-th line decay energy (frequency) for the $i$th species isotope, $P_{ij}$ is $j$-th line decay probability for the $i$-th species isotope, and $V_{\rm Dop}$ is the line Doppler velocity width ($\sim$100--200 \kms). 

The isotropic emission is the local source of \grays\ that eventually travel and scatter through the ejecta. Thus, it only needs to be calculated once before the transfer equation is solved. 
\subsubsection{Scattering Emissivity}
Unlike the prompt emission, the scattering emissivity requires more numerical/computational effort and must be calculated concurrently while solving \eq~\ref{rrte_char_eqn}. The difficulty in calculating the scattering emissivity (\eq~\ref{emissivity_eqn}) is due to the complicated angle and frequency dependence of the anisotropic Klein-Nishina (KN) scattering kernel (\eq~\ref{scatkern}). Since we solve the specific intensity along characteristic rays for all impact parameters $p_i$, we have a fixed grid of polar angles $\theta_i$ (specifically $\mu_i=\cos\theta_i$) for every radius $r_i$ -- note azimuthal symmetry is assumed. The scattering emissivity for an outgoing beam of frequency $\nu'$ and direction $\bf\Omega'$ is generally defined as
	\begin{equation} \label{emissivity_eqn}
		\eta_{\nu'}^{\rm s}(r,{\bf\Omega}') = \int_{0}^{\infty}\frac{\nu'}{\nu} d\nu \oint d{{\bf\Omega}} \sigma_s (\nu \to \nu',\xi) I_\nu(r,\bf{\Omega}),
	\end{equation}
where the prime denotes outgoing, and $\sigma_s (\nu \to \nu',\xi)$ is the KN scattering kernel for a photon scattering with angle given as
\begin{equation}
	\xi={\bf\Omega\cdot\Omega'}=\sqrt{1-\mu^2}\sqrt{1-\mu'^2}\cos(\phi-\phi')+\mu\mu'.
\end{equation} 

Following eq.~7.108 of \cite{Pomraning1973}, the KN scattering kernel for $x=h\nu/m_ec^2$ is given by
	\begin{eqnarray}
		\sigma_s (\nu \to \nu',\xi)&=& N_e \frac{r_o^2}{2} \frac{1}{x \nu} \left[ \frac{x}{x'}+\frac{x'}{x}+2 \left( \frac{1}{x}-\frac{1}{x'} \right) \right. \nonumber \\
	&& \hspace*{-0.35cm} \left. + \left( \frac{1}{x}-\frac{1}{x'} \right) ^2 \right]\delta\left[\xi - \left(1-\frac{1}{x'}+\frac{1}{x}\right)\right]   \label{scatkern} \\
		& \equiv &  N_e\frac{r_o^2}{2}\sigma(\nu,\nu')\delta\left[\xi - \left(1-\frac{1}{x'}+\frac{1}{x}\right)\right]. \qquad \label{scattering_kernel}
	\end{eqnarray}

Given that we assume $I_\nu\neq I_\nu(\phi)$, we need to integrate and remove the $\phi$ dependence in \eq~\ref{emissivity_eqn}. Using the relationship $\delta(f(\phi))=\sum_i \delta(\phi-\phi_i)/|f'(\phi_i)|$ for an arbitrary function $f$ with the zeros $\phi_i$, we can transform our $\delta$-function as
	\begin{equation}
		 \delta\left[ \xi - \left(1-\frac{1}{x'}+\frac{1}{x}\right)\right] \to  \nonumber
	\end{equation}
	\begin{equation}
		\frac{\delta\left[ \phi - \phi_1 \right]}{\left|	\sqrt{ (1-\mu^2)(1-\mu'^2)-(1-\frac{1}{x'}+\frac{1}{x}-\mu\mu')^2 }\right|} \;+ \nonumber
	\end{equation}
	\begin{equation}
		\frac{\delta\left[ \phi - \phi_2 \right]}{\left|	\sqrt{ (1-\mu^2)(1-\mu'^2)-(1-\frac{1}{x'}+\frac{1}{x}-\mu\mu')^2 }\right|},
	\end{equation}	
where
	\begin{eqnarray} \label{zeros}
		\phi_1 &=& \cos^{-1}\left( \frac{1-\frac{1}{x'}+\frac{1}{x}-\mu\mu'}{\sqrt{(1-\mu^2)(1-\mu'^2)}}  \right) + \phi'  \qquad \text{and}\nonumber \\
		\phi_2 &=& 2\pi\,-\,\cos^{-1}\left( \frac{1-\frac{1}{x'}+\frac{1}{x}-\mu\mu'}{\sqrt{(1-\mu^2)(1-\mu'^2)}}  \right) + \phi'.
	\end{eqnarray}
Both $\phi_1$ and $\phi_2$ exist in $\{0,2\pi\}$ since $\cos(\phi-\phi')=\cos(2\pi-[\phi-\phi'])$. Each delta contributes an equal value to the integral (see \eq~\ref{emissivity_eqn}) with respect to $\phi$. Therefore, we have twice the integral of one delta function, giving us a factor of 2. 

This transformation changes the $\mu$ integration limits to make sure the Compton relationship holds.  To find the new $\mu$ limits, we extremize $\xi$ with respect to $\phi$ evaluated at our roots (i.e. $\partial\xi/\partial\phi |_{\phi=\phi_i}=0$). This gives us the constraint that $\phi_i-\phi'=n\pi$, for an integer $n$. Using this result, we find our new limits on $\mu$ to be 
	\begin{equation}\label{limits}
		\mu_{1,2} = \frac{a_1 \pm \sqrt{a_1^2+4a_2}}{2}
	\end{equation}
	\begin{equation}
		\text{for}\qquad a_1 = 2\mu'\left[1+\frac{1}{x}-\frac{1}{x'}\right] \qquad \text{and}\nonumber 
	\end{equation}
	\begin{equation}
 a_2 = \left[\frac{2}{x x'}+\frac{2}{x'} -\frac{1}{x^2} -\frac{2}{x}-\frac{1}{x'^2} -\mu'^2 \right]. \nonumber
	\end{equation}
We can then rewrite \eq~\ref{emissivity_eqn} after integrating over $\phi$ as 
	\begin{equation}
		\eta_{\nu'}^{\rm s}(r,\mu') = N_e r_o^2 \int_{0}^{\infty} \frac{\nu'}{\nu} d\nu \, \sigma(\nu,\nu')\int^{\mu_1}_{\mu_2} d\mu F(\nu,\nu',r,\mu,\mu'), \label{before_cheb} 
	\end{equation}
where
	\begin{equation}
 		F(\nu,\nu',r,\mu,\mu')=\frac{I_\nu(r,\mu)}{\sqrt{(1-\mu^2)(1-\mu'^2)-(1-\frac{1}{x'}+\frac{1}{x}-\mu\mu')^2 }}. \label{mu_integrand}
	\end{equation}
Note that we have cancelled the one half with the factor of 2 from our $\phi$ integration. 
If we look at the $\mu$ integrand in \eq~\ref{before_cheb} with our new $\mu$ integration limits, we run into a singularity at our limits. However, for this integral we can exploit Gauss-Chebyshev quadrature, which is defined as
	\begin{equation}
		\int^1_{-1} \frac{f(x)dx}{\sqrt{1-x^2}}=\sum^n_{i=1}\frac{\pi}{n}f(b_i),
	\end{equation}
for abscissa $b_i=\cos[(2i-1)\pi/2n]$ and integer $n$.
In order to get it into the form of Gauss-Chebyshev quadrature, we can make a linear transformation of $\mu$, namely, $w=c_1\mu+c_2$, for constants $c_1$ and $c_2$. These constants $c_1$ and $c_2$ are determined using the integration limits $\mu_{1,2}$ and solving the following linear equation:
	\begin{equation}
		\left(\begin{array}{cc}
			\mu_1 & 1 \\
			\mu_2 & 1 
		\end{array}\right)
		\left(\begin{array}{c}
			c_1 \\
			c_2
		\end{array}\right)
		=
		\left(\begin{array}{c}
			1 \\
			-1
		\end{array}\right).
	\end{equation}
This has the solution
	\begin{equation}
		\left(\begin{array}{c}
			c_1 \\
			c_2
		\end{array}\right)
		=
		\frac{1}{\mu_1-\mu_2}\left(\begin{array}{c}
			2 \\
			-\mu_1-\mu_2
		\end{array}\right).
		\label{c12}
	\end{equation}

From the definition of $\mu_{1,2}$ in \eq~\ref{limits} and our constants $c_{1,2}$ in \eq~\ref{c12}, we find that our integrand transforms to
	\begin{equation}
		\frac{d\mu}{	\sqrt{(1-\mu^2)(1-\mu'^2)-(1-\frac{1}{x'}+\frac{1}{x}-\mu\mu')^2 }} \to \frac{dw}{\sqrt{1-w^2}}.
	\end{equation}
Finally, \eq~\ref{before_cheb} becomes 
	\begin{eqnarray}
		\eta_{\nu'}^{s}(r,\mu') &=& N_e r_o^2 \int_{0}^{\infty}  \frac{\nu'}{\nu} d\nu \, \sigma(\nu,\nu')\times \nonumber \\
		&& \sum^n_{i=1}\frac{\pi}{n}I_\nu(r,(b_i-c_2)/c_1). \label{eta_after_cheb}
	\end{eqnarray}
This final result for the scattering emissivity is computationally favourable. We avoid having to loop through the large multi-dimensional arrays, thus saving time. 

This transformation has certain limiting cases, such as when the $\mu'=\pm1$. In that case, we can look at the problem two ways. First we have that 
	\begin{eqnarray}
		\lim_{\mu'\to1} (b_i-c_2)/c_1 &=& 1-\frac{1}{x'}+\frac{1}{x} \qquad \text{and} \nonumber \\ 
		\lim_{\mu'\to-1} (b_i-c_2)/c_1 &=& -\left(1-\frac{1}{x'}+\frac{1}{x}\right)
	\end{eqnarray}
Thus, in these cases $I_\nu$ is a constant, and the sum in \eq~\ref{eta_after_cheb} equals $\pi I_\nu(r,\mu=\pm(1-1/x'+1/x))$ -- note there was a factor of 2 from the $\phi$ integration. The second way to understand these cases goes back to \eqs~\ref{emissivity_eqn}~and~\ref{scattering_kernel}. If we look at how the delta function transforms, we have
	\begin{equation}
		 \delta\left[ \xi - \left(1-\frac{1}{x'}+\frac{1}{x}\right)\right] \to \delta\left[ \pm\mu - \left(1-\frac{1}{x'}+\frac{1}{x}\right)\right]
	\end{equation}
In these cases, the $\phi$ integral is $2\pi$, and the $\mu$ integral picks out $\pi I_\nu(r,\mu=\pm(1-1/\gamma'+1/\gamma))$. Both methods produce the same result. 

Our work assumes that there is no contribution from the current frequency to the scattering emissivity (i.e. all photons are down-scattered). This assumption removes coupling between $I_\nu(r,\mu)$ and $\eta_{\nu}^{s}(r,\mu)$ at the current frequency $\nu$, so $I_\nu(r,\mu)$ can be solved exactly at each frequency when formally integrating \eq~\ref{rrte_char_eqn}. 
\subsection{Energy Deposition}
Once we have solved $I_\nu(r,\mu)$ for all depths, we then calculate the energy deposited from scattering at each depth using the following relationship:
	\begin{equation}\label{edep_eqn}
		E_{\rm dep}(r)=E_{\rm lept}(r)+\int_{0}^{\infty}d\nu \oint d\Omega\left[\chi_\nu^{\rm tot}(r)I_\nu(r,\mu)-\eta_{\nu}^{s}(r,\mu)\right],
	\end{equation} 
where $E_{\rm lept}$ is the (assumed) local kinetic energy deposition from decay leptons (positrons and electrons). 
In \eq~\ref{edep_eqn}, the physics we are capturing is the difference between the macroscopic energy lost from the specific intensity and the energy redistributed after scattering. In practice we would only integrate over the range of our frequency grid, which is chosen to cover the physics of the problem.
%
%
\subsection{Implementation}\label{implem}
%
%
As previously mentioned, this code is being implemented as part of \CMFGEN. The code solves \eq~\ref{rrte_char_eqn} along characteristic rays for a given impact parameter ($p_i$) intersecting our radial grid ($r_i$). In this set-up we have $N_{D}$ radial grid points and $N_{C}$ core grid points, making $N_{P}=N_{C}+N_{D}$ impact parameters. 

Since this code treats $\gamma$-rays from radioactive decays, we begin by reading in nuclear decay data such as nuclear decay energies and their decay probabilities for each unstable isotope included in an ejecta model. Lines with decay probabilities $<$1 per cent are not included in this code. However, like \cite{Hillier2012}, we scale the decay line probabilities and decay lepton kinetic energies to conserve the total energy released during decay. Table~\ref{decay_data} lists the following nuclear decay data: half-life, energy per decay, kinetic energies of leptons produced, and line energies and probabilities for the \iso{Ni}{56}$\rightarrow$\iso{Co}{56}$\rightarrow$\iso{Fe}{56} decay chain, which dominates the decay energy for SNe. The annihilation line has a probability of 38 per cent because we assume that each positron produced (with 19 per cent intensity) annihilates without forming ortho-positronium after thermalization. We read in all other supernova data such as the mass fractions of all included species and count either the number of decays since the last time-step (an average) or the instantaneous decay.  
\begin{table}
  \centering
  \begin{tabular}{c c c c}
    \hline\hline
    \multicolumn{4}{c}{ \iso{Ni}{56} $\rightarrow$ \iso{Co}{56} $\rightarrow$ \iso{Fe}{56} } \\ [0.75ex]
    \multicolumn{2}{c}{ \iso{Ni}{56} $\rightarrow$ \iso{Co}{56} } & \multicolumn{2}{c}{ \iso{Co}{56} $\rightarrow$ \iso{Fe}{56} }  \\
    \multicolumn{2}{c}{ $t_{1/2}$ = 6.075 days } & \multicolumn{2}{c}{ $t_{1/2}$ = 77.233 days } \\
    \multicolumn{2}{c}{ $Q_\gamma$ = 1.718 MeV } & \multicolumn{2}{c}{ $Q_\gamma$ = 3.633 MeV } \\
    \multicolumn{2}{c}{ $Q_{\rm th}$ = 0.000 MeV } & \multicolumn{2}{c}{ $Q_{\rm th}$ = 0.116 MeV } \\
    $E_\gamma$ & Prob. & $E_\gamma$ & Prob. \\
    (MeV) &  & (MeV) &  \\
    \hline
    0.158  &  98.8  &  0.511 &  38.0  \\
    0.270  &  36.5  &  0.847 &  100  \\
    0.480  &  36.5  &  0.977 &  1.4  \\
    0.750  &  49.5  &  1.038 &  14.0  \\
    0.812  &  86.0  &  1.175 &  2.3  \\
    1.562  &  14.0  &  1.238 &  67.6  \\
    &  & 1.360 &  4.3  \\
    &  & 1.771 &  15.7  \\
    &  & 2.015 &  3.1  \\
    &  & 2.035 &  7.9  \\
    &  & 2.598 &  17.3  \\
    &  & 3.010 &  1.0  \\
    &  & 3.202 &  3.2  \\
    &  & 3.253 &  7.9  \\
    &  & 3.273 &  1.9  \\
    \hline\hline
  \end{tabular}
  \caption[]{Example nuclear decay data for the \iso{Ni}{56}$\rightarrow$\iso{Co}{56}$\rightarrow$\iso{Fe}{56} decay chain. $t_{1/2}$, $Q_\gamma$, $Q_{\rm th}$ are the half-life, energy per decay, and thermal energy of the leptons produced. We list the decay line energies $E_\gamma$ and probabilities for lines with probabilities $\geq$ 1 per cent. This data and all other nuclear decay data is taken from \url{http://www.nndc.bnl.gov/chart/}.}
  \label{decay_data} 
\end{table}

After reading in all decay line data, we set up a frequency grid that is equally spaced in a log frequency scale for a given regime such as between lines, across the line, and two regimes for the red Compton tail. Each regime's spacing is controlled by input parameters to give a desired spectral resolution. A finer frequency grid will produce ``narrower" spectral line profiles by reducing numerical diffusion in frequency space as we propagate the photons spatially while solving \eq~\ref{rrte_char_eqn}. For a factor of roughly 3 less frequency points, the Gaussian profiles become broader by $\sim$50 per cent with no more than a few percent difference in the energy deposition.

Solving \eq~\ref{rrte_char_eqn} for a given frequency, $k$, introduces difficulty given that the scattering emissivity is an integral over angles at a given depth -- see \eqs~\ref{emissivity_eqn}~and~\ref{scatkern}. No coupling between $\eta_{k}^{\rm s}$ and $I_{k}$ alleviates some computational difficulty. Full calculation of $\eta_{\nu'}^{\rm s}(r,\mu')$ requires integration over all angles for a given depth point, but integration of \eq~\ref{rrte_char_eqn} along rays restricts us to a subset of angles for a given ray -- see Section~\ref{rte_section}. It is necessary to map our intensity and emissivity arrays from $(z,p)\,\rightarrow\,(r,\mu)$ in order to perform all scattering calculations. With our assumption that all scattered photons are downgraded in frequency, we solve \eq~\ref{rrte_char_eqn} exactly from blue to red frequencies. To do this, we calculate the scattering emissivity for all downgraded frequencies from $I_{k}$ and use central differencing quadrature ( $[\nu_{k-1}-\nu_{k+1}]/2$ ) to update $\eta_{j}^{\rm s}$, where $\nu_j<\nu_k$ -- an implicit frequency integration of \eq~\ref{eta_after_cheb}. 

We interpolate $I_{k}$ (using monotonic cubic interpolation) onto a finer equally spaced linear $\mu$ grid in order to use Gauss-Chebyshev abscissa ($b_i=\cos\left[(2i-1)\pi/2n\right]$). An equally spaced linear grid allows us to quickly find and select the angle abscissa. After we update $\eta_{\nu'}^{\rm s}$ for all possible down-scattered frequencies from $\nu_k$, we then map our arrays back into $(z,p)$ and solve for $I_{k+1}$ for all $p$. 

In principle, the time required to calculate the scattering emissivity scales as $N_D\times N_\nu^2 \times N_\mu^2$, where $N_\nu$ is the number of frequency points and $N_\mu$ is the maximum number of angle points equal to $2N_P-1$. However, we loop over down-scattered frequencies when calculating $\eta_{\nu'}^{\rm s}(r,\mu')$, so calculation time will scale less than $N_D\times N_\nu^2 \times N_\mu^2$. Using Gauss-Chebyshev quadrature replaces one loop of length $N_\mu$ for a loop of length $N_{\rm Cheb}$. Interpolation onto a monotonic $\mu$ grid circumvents looping to find the abscissa in our arrays and makes its calculation time tractable. 

Once we have calculated $I_\nu(z,p)$ for all frequencies and impact rays, we  map it back into the $(r,\mu)$ space and from \eq~\ref{edep_eqn} calculate the energy deposited from \grays. This decay energy deposition will then be used and read in by \CMFGEN\ as a non-thermal heating source when solving the rate equations coupled to the \rrte\ for lower energy frequencies. We calculate an observer's frame flux according to \cite{Hillier2012}, which can be compared to observed \gray\ spectra of SNe. 

\begin{figure}
\centering
\hspace*{-0.5cm}
\includegraphics[scale=0.87]{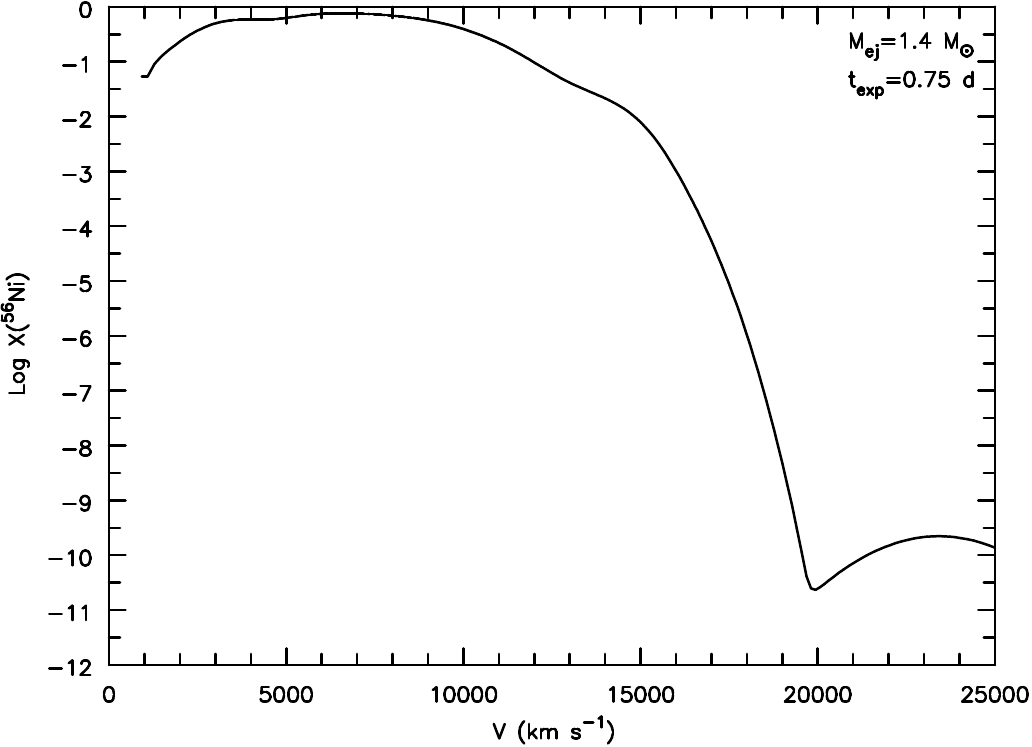}
\vspace*{-0.5cm}
\caption[]{\iso{Ni}{56} mass fraction at 0.75 days after the explosion for the \Mch\ ejecta (CHAN) model of \cite{Wilk2018}.}
\label{xmass_nick}
\end{figure}

\begin{figure*}
\begin{minipage}[t]{0.5\linewidth}
\centering
\hspace*{-4.75cm}\includegraphics[scale=0.88]{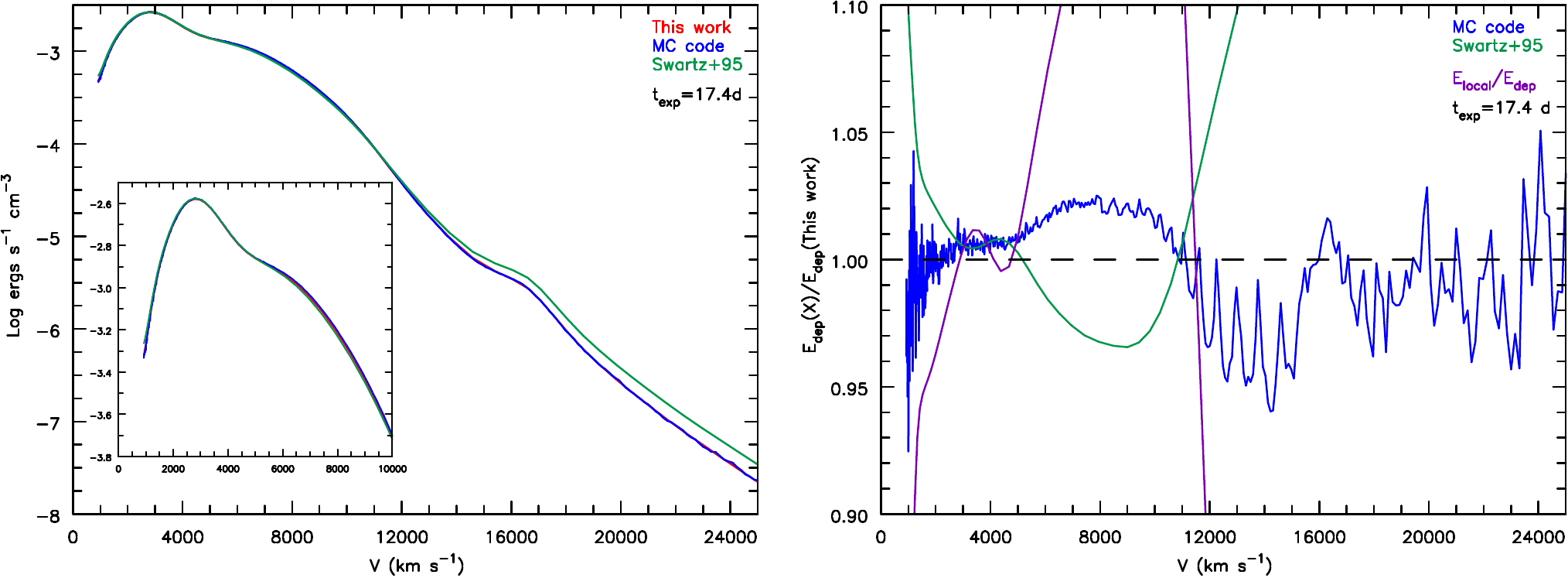}
\end{minipage}
\vspace*{-0.2cm}
\caption[]{Comparison between this work, the MC method by \cite{Hillier2012}, and \cite{Swartz1995} of the energy deposited by both leptons and $\gamma$-rays from nuclear decays at 17.4 days post-explosion in a Chandrasekhar mass WD with 0.62 \Msun\ of initial \iso{Ni}{56}. The MC method and the method described in this work agree within 3 per cent despite fundamental differences in their approach. Discrepancies in the inner region result from MC statistical effects from little mass in the inner region. Shown in purple is the ratio of the local energy emitted to the energy deposited.}
\label{edep_1}
\end{figure*}

\begin{figure*}
\begin{minipage}[t]{\linewidth}
\centering
\hspace*{-0.5cm}\includegraphics[scale=0.88]{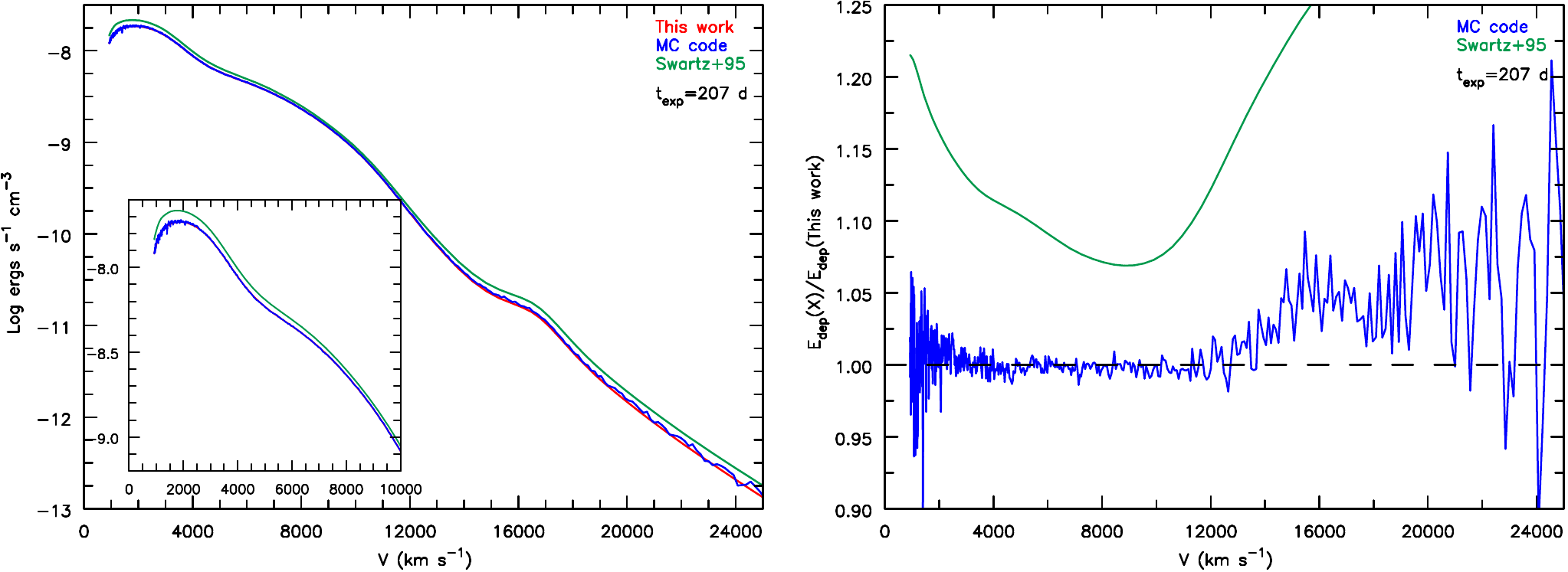}
\end{minipage}
\vspace*{-0.4cm}
\caption[]{Comparison between this work, the MC method by \cite{Hillier2012}, and \cite{Swartz1995} of the energy deposited by both leptons and $\gamma$-rays from nuclear decays at 207 days post-explosion in a Chandrasekhar mass WD with 0.62 \Msun\ of initial \iso{Ni}{56}. The MC method and the method described in this work agree within $\sim$1 per cent despite fundamental differences in their approach. Discrepancies in the inner region result from MC statistical effects from little mass in the inner region.}
\label{edep}
\end{figure*}
\begin{figure*}
\begin{minipage}[t]{\linewidth}
\centering
\hspace*{-0.5cm}
\includegraphics[scale=0.85]{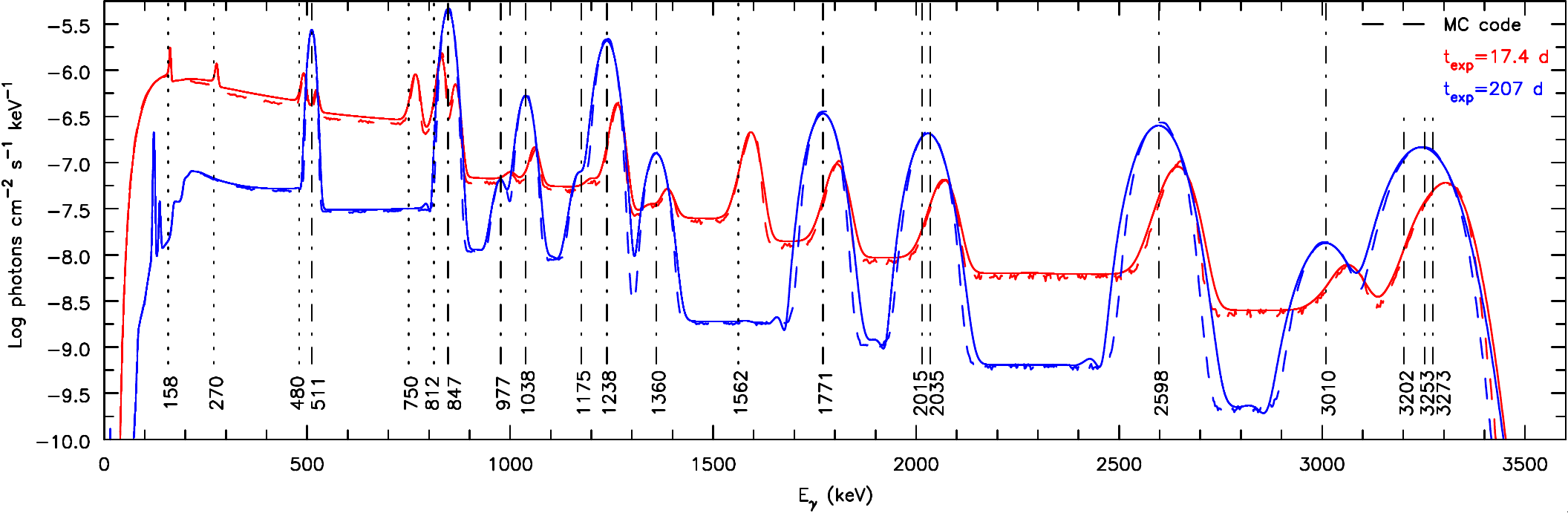}
\end{minipage}
\vspace*{-0.4cm}
\caption[]{Synthetic \gray\ spectra at two different epochs--17.4 and 207 days post-explosion in a Chandrasekhar mass WD with 0.62 \Msun\ of initial \iso{Ni}{56}. Flux counts are relative to a distance of 3.5 Mpc in comparison to SN 2014J in M82 \citep{Karachentsev2006}. Dotted lines correspond to the flux calculated by the MC code from the appendix of \cite{Hillier2012}.
}
\label{spectra}
\end{figure*}
%
%
\section{Results}\label{results}
%
%
For this work, we recomputed model CHAN, a Chandrasekhar mass (\Mch) WD with 0.62 \Msun\ of \iso{Ni}{56} initially, from \cite{Wilk2018} at two epochs, 17.4 days after explosion (roughly bolometric maximum) and 207 days after explosion (nebular time -- optically thin to \grays). The initial \iso{Ni}{56} mass fraction at 0.75 d is shown in \fig~\ref{xmass_nick} for model CHAN. We performed the calculations described in this work and compared the results to two other methods \CMFGEN\ can use to calculate the energy deposition: (1) MC transport for \grays\ \citep{Hillier2012} using 8\,000\,000 decays and (2) a grey absorption approximation \citep{Swartz1995} using $\kappa_\gamma=0.06Y_e$ cm$^2$ g$^{-1}$.

\subsection{Runtime}\label{results_timing}
%

The runtime scaling of our \gray\ transport code with the number of depth points, number of angles, and the number of frequency points is
 explained in Section~\ref{implem}. To improve efficiency we have made this code parallelizable over depth and have tested it using an Intel(\textregistered) Xeon(\textregistered) CPU E5-4610 2.40GHz processor and 8 cores. Tests on more modern processors (like Intel(\textregistered) Xeon(\textregistered) CPU E5-2620 v4 2.10GHz) show an improvement of a factor of two in speed (for the same number of cores). All calculations were performed with $N_D=109,\;N_C=15, and \;N_P=124$.  A calculation with a very fine frequency grid resolution ($\sim$26\,600 frequency points), needed if very accurate line profiles are to be computed, has a runtime of $\sim$17 hours. However, for most work we are only interested in the energy deposition rate, and we can use a much lower spectral resolution. For example, for a low spectral resolution calculation with $\sim$6\,500 frequency grid points the code's runtime is approximately 45 minutes (the runtime scales roughly as the number of frequency points squared). Even though the number of frequency points has been reduced by a factor of $\sim$4, the energy deposited throughout the ejecta differs by at most 1 per~cent from the high resolution calculation. 


The runtime on the same machine for a MC calculation on a single processor with 8\,000\,000 decays (necessary for low statistical noise) is significantly longer than our low spectral resolution calculation. In this case, the MC runtime is roughly 10 hours. Using a factor of 10 less decays per species, the MC calculation runtime is roughly 1 hour. For lower resolution calculations, the runtimes of both codes are somewhat comparable. 
\subsection{Energy Deposition}\label{results_edep}
%
\fig~\ref{edep_1} compares the ratio of the energy deposition at 17.4 days calculated using our new radiative transfer code to that computed with the MC code, and to that obtained using the grey absorption approximation, as a function of velocity for a \Mch\ WD.
\fig~\ref{edep_1} shows that our work is in agreement with the MC method within 3 per cent at $<$20\,000 \kms. Below 3000 \kms, the MC method is subject to statistical noise and has discrepancies with this work due to a ``\iso{Ni}{56} hole" where little radioactive material is mixed in. 
Beyond 20\,000 \kms, MC statistical noise is the source of the discrepancy between the two codes. The error between 5\,000 and 11\,000 \kms\ is partially numerical since doubling the value of $N_D$ caused the error in this region to decrease. However, doubling the value of $N_\nu$ gave minimal improvement. The error is also likely sensitive to the interpolation techniques. However, despite our best efforts, we were unable to reduce the discrepancy below 1\,per~cent.

Table~\ref{table_edep} lists the total integrated energy deposition over the whole ejecta at this epoch and shows that the two methods agree within $\sim$2.5 per cent. \fig~\ref{edep_1} also shows the ratio of the non-thermal energy deposited to that of local energy released from nuclear decays. We see that beyond 12\,000 \kms, the energy deposition comes from the inner ejecta as the \gray\ photons scatter. In the region between 12\,000--20\,000 \kms\ where many optical and diagnostic lines are formed, this work is consistent within 2.5--3 per cent to that of \cite{Hillier2012}. In the same region, the energy deposition computed using the grey approximation diverges from the other methods. 

\fig~\ref{edep} is the same as \fig~\ref{edep_1}, except now at 207~days and without the ratio of the energy deposition to the local energy emitted being plotted. Despite the fundamental differences in the approach each code uses to calculate the energy deposition, the MC code and our work agree to within 1 per cent (highlighted in Table~\ref{table_edep}). At late times and for ejecta velocities less than 10\,000 \kms, the grey approximation is inconsistent with the two other methods by more than $\sim$5 per cent. At this epoch, important strong cooling lines form at velocities $\leq$10\,000 \kms, so  wrongly estimating the energy deposited may affect the ionization structure and/or flux in strong cooling lines. 
\begin{table*}
 \centering 
 \begin{tabular}{|c|c|c|c|c|c|c|}
  \hline
                    & \multicolumn{3}{c}{$t_{\rm exp}$ = 17.4 days} & \multicolumn{3}{c}{$t_{\rm exp}$ = 207.0 days}   \\ [0.75ex]
                    & $E_{\rm dep}$ (\ergsec)   & $L_{\rm escape}$ (\ergsec) & $L_{\rm flux}$ (\ergsec) & $E_{\rm dep}$ (\ergsec) &  $L_{\rm escape}$ (\ergsec) & $L_{\rm flux}$ (\ergsec)  \\ \hline
  This work         &  1.260(43) & 9.669(41) & 8.466(41) & 9.273(40) & 1.343(42) & 1.359(42) \\
 \cite{Hillier2012} &  1.279(43) & 7.892(41) & 7.762(41) & 9.291(40) & 1.343(42) & 1.305(42) \\
  \hline
 \end{tabular}
 \caption{Listed is the total energy deposition integrated over the whole ejecta ($E_{\rm dep}$) and the integrated flux from the synthetic spectrum ($L_{\rm flux}$).}
\label{table_edep}
\end{table*}
%
\subsection{Synthetic Spectra}\label{results_spectra}
%
From the CMF at the outer boundary, we can transform the specific intensity into the observer's frame to produce a synthetic \gray\ spectrum -- see section 11 of \cite{Hillier2012}. \fig~\ref{spectra} shows our resulting synthetic spectra calculated at two epochs, 17.4 and 207 days post-explosion. At 17.4 days, the dominant decay luminosity begins to switch from \iso{Ni}{56} to \iso{Co}{56}, and the spectrum shows strong lines from both \iso{Ni}{56} and \iso{Co}{56} -- see Table~\ref{decay_data}. However, at 207 days, all the \iso{Ni}{56} has decayed and the synthetic spectra is dominated by \iso{Co}{56} decay lines. At both epochs, synthetic spectra from our work and the MC method are in good agreement. The total integrated flux listed in Table~\ref{table_edep} shows that the two methods are within $\sim$9 per cent at 17.4 days and $\sim$4 per cent at 207 days.

\begin{figure}
\begin{minipage}[t]{\linewidth}
\centering
\hspace*{-0.5cm}
\includegraphics[scale=0.87]{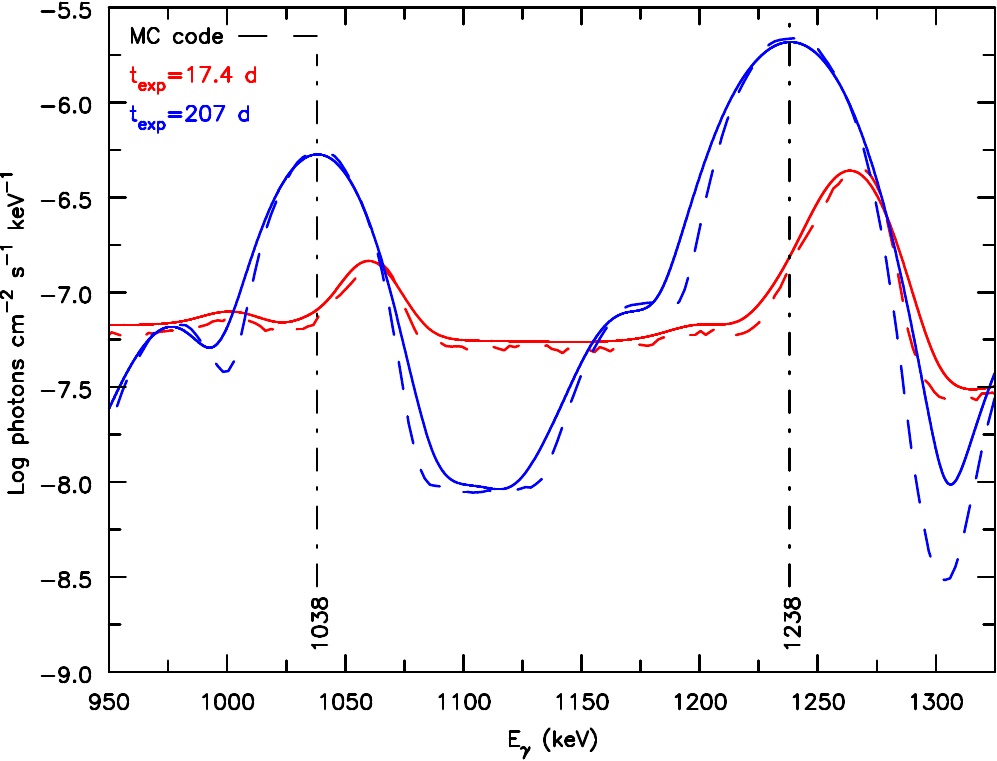}
\end{minipage}
\vspace*{-0.4cm}
\caption[]{Synthetic flux same as \fig~\ref{spectra}, but we have added vertical lines at line centre energy \iso{Co}{56} 1038 and 1238 keV. Since the red side of the line has a larger optical depth compared to the blue, we see stronger emission on the blue side of the line profile} when the optical depth is high at early times.
\label{asymmetry}
\end{figure}

Both the MC method and our radiative transfer code produce synthetic spectra with predicted asymmetric profiles with absorption on the red side of the emission line. These asymmetric profiles are not uncommon. They are predicted and seen in X-ray line profiles for massive stars \citep{Macfarlane1991,OwockiCohen2001,Cohen2010,Cohen2014}. They are also a product of dust scattering \citep{Romanik1981}, and have been modelled for dust in the ejecta of SN1987A \citep{Bevan2016}. Electron scattering opacity also produces blue shifted asymmetric profiles for some optical lines in Type II SNe \citep{DessartHillier2005}. Many previous theoretical studies have predicted the anticipated asymmetric \gray\ line profiles, notably \cite{Burrows1990,Mueller1991,Hoeflich1992,Hoeflich1993,Hoeflich1994,Maeda2006}. Since Compton scattering is a continuum opacity, the optical depth of the red side of the line is higher because the path length is larger to the far side of the ejecta. We expect our profiles to exhibit the same effect. In \fig~\ref{asymmetry} we highlight two \iso{Co}{56} decay lines at 1038 and 1238 keV. \fig~\ref{asymmetry} shows that at 17.4 days our profiles are asymmetric as the optical depth to \grays\ is large, causing most of the emission to be in the blue side of the line profile; whereas, at 207 days its optical depth is low, and the profile is symmetric.

As can be seen from \figs~\ref{spectra} and \ref{asymmetry}, the profiles produced by the MC calculation are somewhat narrower than those produced by our \gray\ transfer approach. This arises from numerical diffusion as we propagate photons from the inner regions to the outer boundary of the model (as the calculation is done in the CMF, the photons are propagated in both frequency and space). Numerical diffusion can be reduced by increasing the grid resolution, or by reducing the extent of the outer boundary (especially relevant at 207 days). The best approach would be to utilise the computed scattering emissivities in an observer's frame calculation, but given the lack of high  quality observed data we have not implemented such a procedure.

%
\subsection{Comparison to SN2014J}\label{sect_sn2014J}
\begin{figure}
\begin{minipage}[t]{\linewidth}
\centering
\hspace*{-0.5cm}
\includegraphics[scale=0.75]{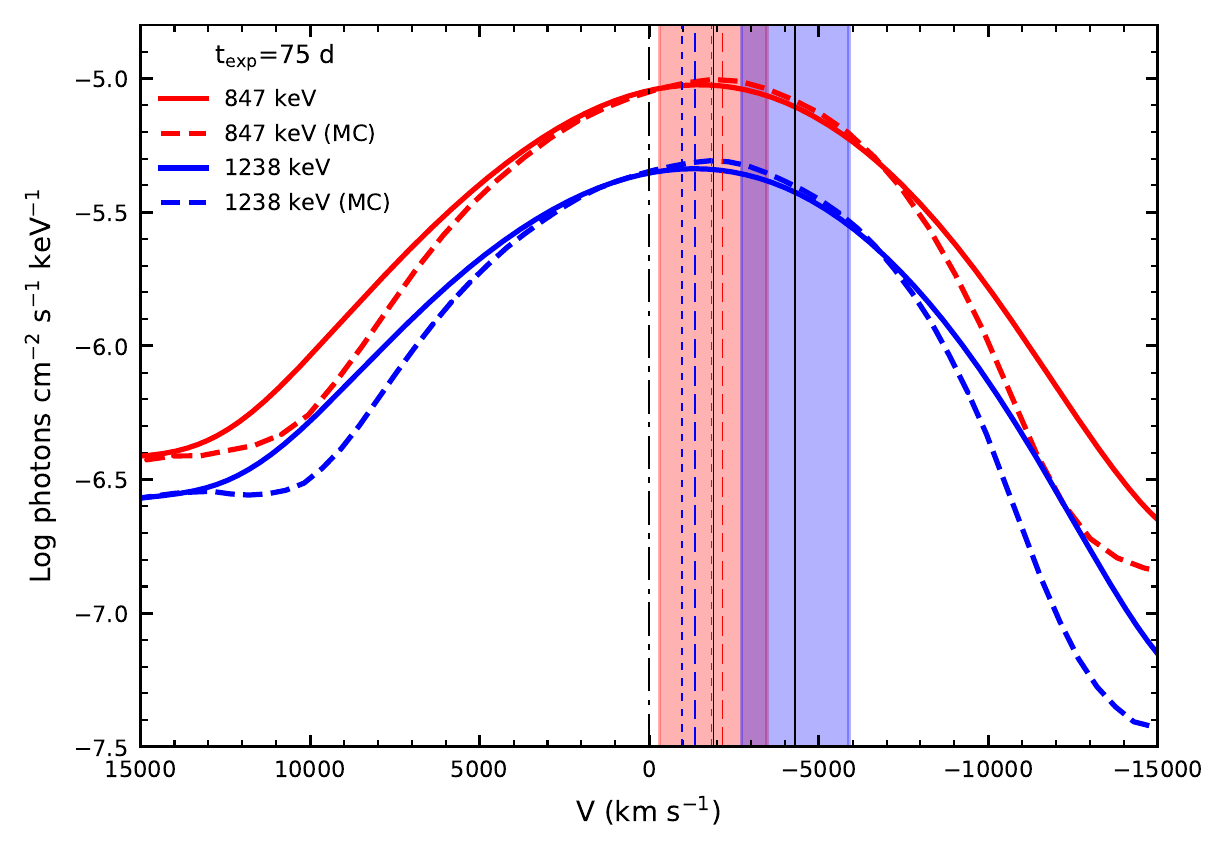}
\end{minipage}
\vspace*{-0.5cm}
\caption[]{Synthetic flux line velocities for 847 keV and 1238 keV computed from our ejecta model at 75 days post-explosion. The 847 keV line centroid velocities are -1836 and -2158 \kms\ for this work and the MC code respectively. Also, the 1238 keV line centroid velocities are -960 and -1348 \kms\ respectively. We have added a vertical lines at 0 \kms\ (dot-dashed), -1836 \kms\ (red dotted), -2158 \kms\ (red dashed), -960 \kms\ (blue dotted), -1348 \kms\ (blue dashed), -1900 \kms\ (solid black), and -4300 (solid black). The red shaded region represents the 1$\sigma$ 1600 \kms\ uncertainty from -1900 \kms, given the data from SN2014J for the 847 keV line. The blue shaded region represents the 1$\sigma$ 1600 \kms\ uncertainty from -4300 \kms, given the data from SN2014J for the 1238 keV line.}
\label{velocity}
\end{figure}

%
To compare our work to the observations from SN 2014J, we computed \gray\ synthetic spectra using our ejecta model at 75 days. \fig~\ref{velocity} shows the 847 and 1238 keV line profiles as a function of velocity, comparing synthetic spectra computed from this work and the MC code at 75 days. The 847 keV line centroid velocities are $-$1836 and $-$2158 \kms\ for this work and the MC method respectively. Similarly, the 1238 keV line centroid velocities are $-$960 and $-$1348 \kms. These 847 keV line results are consistent with the values measured for the \gray\ spectrum obtained for SN2014J. \citet[][in fig.~4 and table~1]{Churazov2014} shows that the 847 keV cobalt line is slightly blueshifted with a velocity of $-$1900$\pm$1600 \kms. However, our work disagrees with the 1238 keV cobalt line. Table~1 of \cite{Churazov2014} shows this line to have a peak velocity shift of $-$4300$\pm$1600 \kms. This line should have a smaller blueshifted velocity relative to the 847 keV line since the optical depth is lower due to the smaller cross section at 1238 keV. Given the very large errors on the mean shifts, the disagreement may simply be statistical. 
The fiducial model plotted in fig~4 of \cite{Churazov2014} shows a less blueshifted 1238 keV line profile.

Not only do our profiles agree with those measured by \cite{Churazov2014}, but our flux levels also agree. Adjusting our flux at 75 d in \fig~\ref{velocity} for a distance of 3.5 Mpc to M82, our flux levels are roughly $\sim9\times10^{-6}$ photons cm$^{-2}$ s$^{-1}$ keV$^{-1}$ for our 847 keV line, consistent with the flux levels shown in fig~1 of \cite{Churazov2014}.
%
%
%
\subsection{Grey transfer}\label{sect_grey_transfer}
%
%
Since SN radiative transfer calculations are already time-intensive, it is beneficial to use a simple and fast prescription to calculate the energy deposited by \grays\ in the ejecta. The grey absorption method of \cite{Swartz1995} (hereafter S95) is one such fast procedure to calculate the energy deposition. Comparing the results of both the MC radiative transfer and this work from \figs~\ref{edep_1}~and~\ref{edep}, we see that the grey approximation of S95 would require a time varying grey opacity factor as well as one that varies spatially. Simply using the mass absorption coefficient, $\kappa_\gamma=0.06Y_e$ cm$^2$ g$^{-1}$, does not reproduce the energy deposition the other methods produce. 

\begin{figure*}
\begin{minipage}[t]{\linewidth}
\centering
\hspace*{-0.2cm}
\includegraphics[scale=1.05]{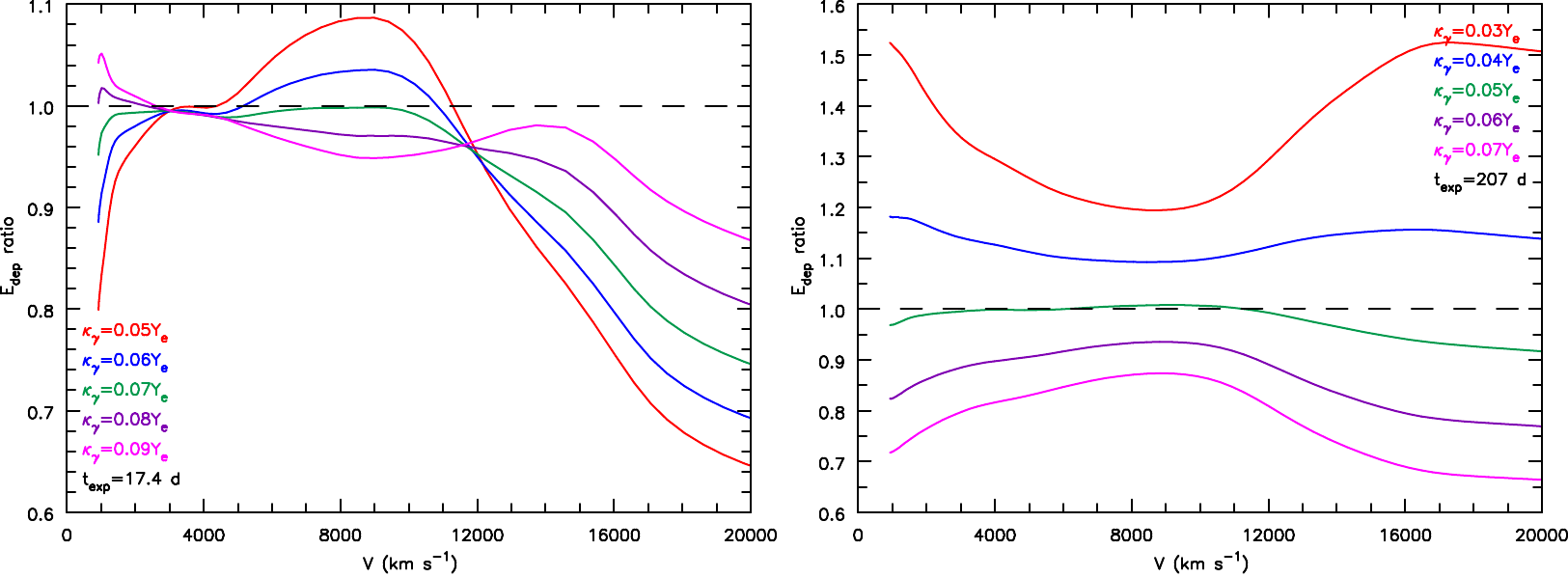}
\end{minipage}
\vspace*{-0.4cm}
\caption[]{Energy deposition ratio comparison of grey radiative transfer calculations of \cite{Swartz1995} using $\kappa_\gamma=\alpha Y_e$ cm$^2$ g$^{-1}$ ($\alpha$ = 0.03, 0.04, 0.05, 0.06, 0.07, 0.08, 0.09). This ratio corresponds to $E_{\rm dep}$(this work)$/E_{\rm dep}$(grey).}
\label{Grey_comparison}
\end{figure*}

\fig~\ref{Grey_comparison} shows the ratio of the calculated energy deposition of this work to that calculated using the grey transfer from S95. However, we show the energy deposition ratio for varying coefficients for $\kappa_\gamma$ at 17.4 and 207 days in the grey approximation. At 17.4 days, we see that $\kappa_\gamma=0.07 Y_e$ cm$^2$ g$^{-1}$ matches to our work below 10\,000 \kms, while $\kappa_\gamma=0.09 Y_e$ cm$^2$ g$^{-1}$ more accurately reproduces the energy deposition beyond 10\,000 \kms. For low values of the grey absorption (i.e. 0.05$Y_e$ cm$^2$ g$^{-1}$) too little energy is deposited in the inner ejecta, which is instead deposited in the outer region. However, increasing the constant in the grey absorption coefficient still produces too much absorption in the outer ejecta.

At nebular times, we see that a lower value of $\kappa_\gamma=0.05Y_e$ cm$^2$ g$^{-1}$ reproduces the energy deposition of the other methods when the ejecta is optically thin to \grays. However, \fig~\ref{Grey_comparison} shows too much energy being deposited into the outer ejecta beyond 12\,000 \kms\ at nebular times. More sophisticated approaches like that of \cite{Jeffery1998} may be needed to model different parts of the ejecta as a function of time.

These values of $\kappa_\gamma$ required to reproduce the energy deposition are roughly aligned with those of \cite{Maeda2006}. \cite{Maeda2006} argue that $\kappa_\gamma=0.027$ cm$^2$ g$^{-1}$ best reproduces a light curve of their spherically symmetric F model. With $Y_e\approx 0.5$, the result of \cite{Maeda2006} is consistent with the value $\kappa_\gamma=0.05Y_e$ cm$^2$ g$^{-1}$ that we claim agrees with our nebular energy deposition. However, our work demonstrates that values of much higher $\kappa_\gamma$ are required to reproduce the energy deposition at early times (see \fig~\ref{Grey_comparison}). \cite{Woosley1986} claim a higher value (because of the centrally located distribution of \iso{Ni}{56} and sensitivity to angle averaging effects along density gradients) of $\kappa_\gamma=0.07$ cm$^2$ g$^{-1}$ reproduces the energy deposition function.
%
%
\section{Conclusion}\label{conclusion}
%
%
We have presented a new code that solves the \rrte\ for \grays, taking into account opacity, prompt radioactive decay emissivity, and scattering emissivity. In computing the scattering emissivity, we assume that all photons are downgraded in energy and ignore any thermal redistribution effects since the expansion velocities dominate the transfer. From the specific intensity, we are able to produce an observer's frame spectrum as well as the energy deposition consistent with that of the MC code of \cite{Hillier2012}. 

For a low spectral resolution ($\sim$6\,500 frequency grid points) calculation, our new code has the advantage of running in approximately 45 minutes using parallel processing with 8 CPUs. Low resolution calculations result in at most 1 per cent differences in calculated energy deposition within the ejecta compared to the higher spectral resolution. In comparison to the MC code, with 8\,000\,000 decays needed to achieve low statistical noise, the code runs in approximately 10 hours on the same machine using one CPU. In terms of the integrated energy deposition, the two codes agree within 3 per cent at early times and within 1 per cent at late times.

We have shown that this code produces the expected line profiles. When the optical depth to \grays\ is large, the red side of the line has a higher optical depth than the blue side, and thus most of the emission comes out in the blue side of the line profile -- see \figs~\ref{spectra}, \ref{asymmetry}, \ref{velocity}, and \citet[][fig.~4 and table~1]{Churazov2014}.

This code will be publicly available and serves (along with all other MC \gray\ radiative transfer codes) to improve the astrophysics community's constraints on nucleosynthetic yields as well as the stratification of nuclear material in SN ejecta. We are currently limited by observations of \grays\ from SNe, so future observations of \grays\ will uncover a previously untapped opportunity to understand more about the nature of SNe and their progenitors.

%
\section*{acknowledgements}
%
DJH acknowledges partial support from STScI theory grant HST-AR-12640.001-A, and DJH and KDW thank NASA for partial support through theory grant NNX14AB41G.
\label{lastpage}
\input{Gammaray_code_biblio_v2.bbl}

\end{document}